\documentclass{article}
\usepackage{spconf}
\usepackage{hyperref}
\usepackage{url}
\usepackage[utf8]{inputenc}
\usepackage{graphicx}
\usepackage{cite}
\usepackage{color}
\usepackage{tabularx}
\usepackage{soul}
\usepackage{boldline}
\usepackage{multirow}

\title{Multitask learning for frame-level instrument recognition}
\name{Yun-Ning Hung$^1$, Yi-An Chen$^
2$ and Yi-Hsuan Yang$^1$}
\address{$^{1}$ Research Center for IT Innovation, Academia Sinica, Taiwan \\
$^{2}$ KKBOX Inc., Taiwan\\
\tt \{biboamy,yang\}@citi.sinica.edu.tw, annchen@kkbox.com}

\begin{document}

\maketitle
\begin{abstract}
For many music analysis problems, we need to know the presence of instruments for each time frame in a multi-instrument musical piece.
However, such a frame-level instrument recognition task remains difficult, mainly due to the lack of labeled datasets. 
To address this issue, we present in this paper a large-scale dataset that contains synthetic polyphonic music with frame-level pitch and instrument labels. 
Moreover, we propose a simple yet novel network architecture to jointly predict the pitch and instrument  for each frame.
With this multitask learning method, the pitch information  can be leveraged to predict the instruments, and also the other way around. 
And, by using the so-called pianoroll representation of music as the main target output of the model, our model also predicts the instruments that play each individual note event. 
We validate the effectiveness of the proposed method for frame-level instrument recognition by comparing it with its single-task ablated versions and three state-of-the-art methods.
We also demonstrate the result of the proposed method for multi-pitch streaming with real-world music. For reproducibility, we will share the code to crawl the data and to implement the proposed model at: \href{https://github.com/biboamy/instrument-streaming/}{https://github.com/biboamy/instrument-streaming/}.
\end{abstract}


\section{Introduction}
Pitch and timbre are two fundamental properties of musical sounds. While the pitch decides the notes sequence 
of a musical piece, the timbre decides the instruments used to play each note.
Since music is an art of time, for detailed analysis and modeling of the information of a musical piece, we need to build a computational model that predicts the pitch and instrument labels for each time frame. 
With the release of several datasets \cite{bittner2014medleydb, thickstun2017learning} and the development of deep learning techniques, recent years have witnessed great progress in frame-level pitch recognition, a.k.a., multi-pitch estimation (MPE) \cite{Bittner2017DeepSR, Thickstun2018InvariancesAD}. However, this is not the case for  the instrument part, presumably due to the following two reasons.

\begin{figure}[t]
\centering
\includegraphics[width=0.48\textwidth]{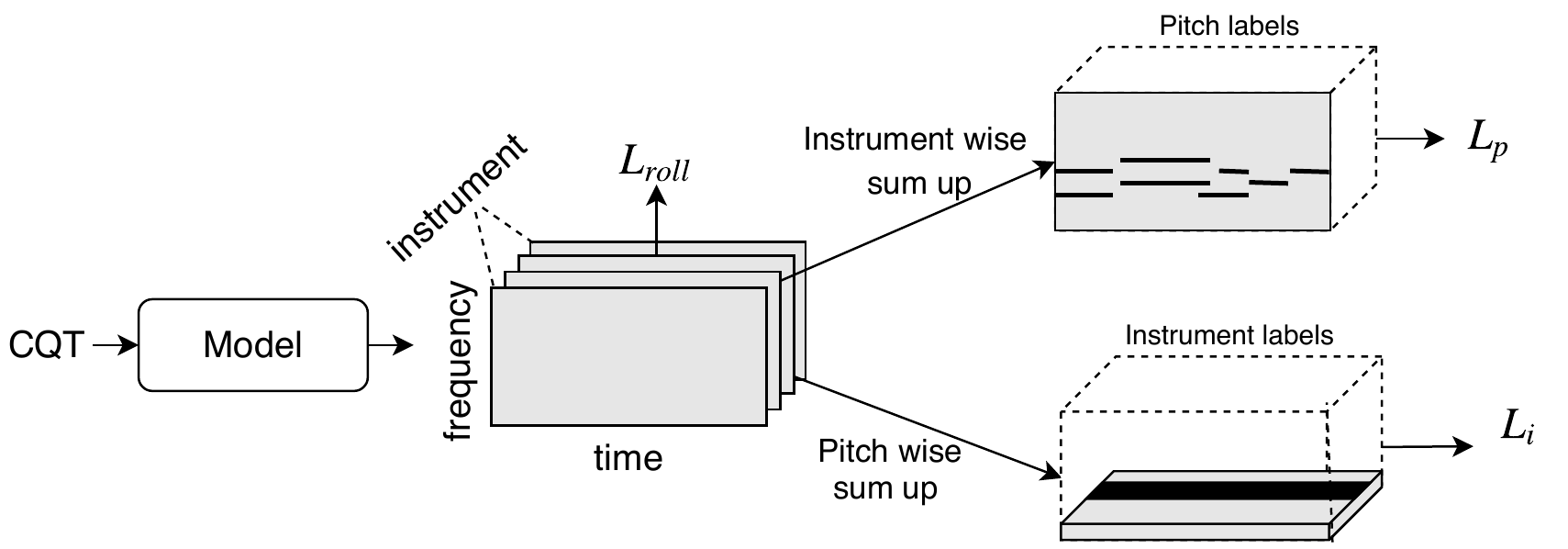}
\caption{Architecture of the proposed model, which employs three loss functions for predicting the (multitrack) pianoroll, the pitch roll, and the instrument roll.
The pitch and instrument
predictions are computed directly from the predicted pianoroll, which is a tensor of \{frequency, time, instrument\}.}
\label{fig: streaming}
\end{figure}

First, manually annotating the presence of instruments for each time frame in a multi-instrument musical piece is a time-consuming and labor-intensive process. As a result, most datasets available to the public only provide instrument labels on the \emph{clip level}, namely, labeling which instruments are present over an entire audio clip of possibly multi-second long \cite{joder09taslp, bosch12ismir,openmic,AUDIOSET}. Such clip-level labels do not specify the presence of instruments for each short-time frame (e.g., multiple milliseconds, or for each second).
Datasets with frame-level instrument labels emerge only over the recent few years \cite{thickstun2017learning,bittner2014medleydb,Duan2010MultipleFF,Gururani2017MixingS}. However,
as listed in Table \ref{tab:dataset} (and will be discussed at length in Section \ref{sec:background}), these datasets contain at most a few hundred songs and some of them contain only classical musical pieces. The musical diversity found in these datasets might therefore not be sufficient to train a deep learning model that performs well for different musical pieces. 


\begin{table*}[t]
\centering
\begin{tabular}{|r|l l l l l|} \hline
& Pitch labels & Instrument labels & Real or Synth & Genre & Number of Songs\\
 \hline\hline
 MedleyDB \cite{bittner2014medleydb}& $\triangle$
\cite{Bittner2017DeepSR,kimSLB18} & ${\surd}$ \cite{Li2015AutomaticIR,Gururani2018InstrumentAD} & Real & Variety & 122\\
MusicNet \cite{thickstun2017learning} & ${\surd}$ \cite{Thickstun2018InvariancesAD} & ${\surd}$ \cite{Hung2018FramelevelIR} & Real & Classical & 330\\
Bach10 \cite{Duan2010MultipleFF}& ${\surd}$ \cite{Duan2010MultipleFF} & ${\surd}$ \cite{Giannoulis2014ImprovingIR} & Real & Classical & 10\\
Mixing Secret \cite{Gururani2017MixingS}&  & ${\surd}$ \cite{Gururani2018InstrumentAD} & Real & Variety & 258\\
MuseScore (this paper) & ${\surd}$ & ${\surd}$ & Synthetic & Variety & 344,166\\
\hline
\end{tabular}
\caption{This table provides information regarding some datasets that provide frame-level labels for either pitch or instrument: whether the audio is real or synthetic, the genre and the number of  songs. We also cite some papers (after the symbols $\surd$ or $\triangle$) that employed these datasets for training either pitch or instrument recognition models. And, we use $\triangle$ to denote `part of it.'}
\label{tab:dataset}
\end{table*}

Second, we note that most recent work that explores deep learning techniques for frame-level instrument recognition focuses only on the instrument recognition task itself and adopts the \emph{single-task} learning paradigm \cite{liu2018weakly,Gururani2018InstrumentAD,Hung2018FramelevelIR}. 
This has the drawback of neglecting the strong relations between pitch and instruments. 
For example, different instruments have their own pitch ranges and tend to play different parts in a polyphonic musical composition. 
Proper modeling of the onset and offset of musical notes may also make it easier to detect the presence of instruments \cite{Hung2018FramelevelIR}. 
From a methodological point of view, we see a potential gain to do better than these prior arts by using a \emph{multitask} learning paradigm that models timbre and pitch jointly.
This requires a dataset that contains both frame-level pitch and instrument labels.


In this paper, we introduce a new large-scale dataset called \emph{MuseScore} to address these needs. 
The dataset contains the audio and MIDI pairs for 344,166 musical pieces downloaded from the official website (\url{https://musescore.org/}) of MuseScore, an open source and free music notation software licensed under GPL v2.0.
The audio is synthesized from the corresponding MIDI file, usually using the sound font of the MuseScore synthesizer. Therefore, it is not difficult to temporally align the audio and MIDI files to get the frame-level pitch and instrument labels for the audio.
Although the dataset only contains synthesized audio, it includes a variety of performing styles in different musical genres. 


Moreover, we propose to transform each MIDI file to the \emph{multitrack pianoroll} representation of music (see Fig. \ref{fig: streaming} for an illustration) \cite{pypianoroll}, which is a binary tensor representing the presence of notes over different time steps for each instrument.
Then, we propose a multitask learning method that learns to predict 
from the audio of a musical piece
its (multitrack) pianoroll, frame-level pitch labels (a.k.a., the \emph{pitch roll}), and the instrument labels (a.k.a., the \emph{instrument roll}).
While the latter two can be obtained by directly summing up the pianoroll along different dimensions, the three involved loss functions would work together to force the  model learn the interactions between pitch and timbre. 
Our experiments show that the proposed model can not only perform better than its task-specific counterparts, but also existing methods for frame-level instrument recognition \cite{liu2018weakly,Gururani2018InstrumentAD,Hung2018FramelevelIR}.



\section{Background}
\label{sec:background}


To our knowledge, there are four public-domain datasets that provide frame-level instrument labels, as listed in Table \ref{tab:dataset}. 
Among them, MedleyDB \cite{bittner2014medleydb}, MusicNet \cite{thickstun2017learning} and Bach10  \cite{Duan2010MultipleFF} are collected originally for MPE research, while Mixing Secret \cite{Gururani2017MixingS} is meant for instrument recognition. 
When it comes to building ``clip-level'' instrument recognizers, there are other more well-known datasets such as the ParisTech \cite{joder09taslp} and IRMAS \cite{bosch12ismir} datasets. Still, there are previous work that uses these datasets for building either clip-level \cite{Li2015AutomaticIR,Giannoulis2014ImprovingIR} or frame-level \cite{Hung2018FramelevelIR,Gururani2018InstrumentAD} instrument recognizers.

There are three recent works on frame-level instrument recognition. 
The model proposed by Hung and Yang \cite{Hung2018FramelevelIR} is trained and evaluated on different subsets of MusicNet \cite{thickstun2017learning}, which consists of only classical music.
This model considers the pitch labels estimated by a pre-trained model (i.e. \cite{Bittner2017DeepSR}) as an additional input to predict instrument, but the pre-trained model is fixed and not further updated. 
The model presented by Gururani \emph{et al.} \cite{Gururani2018InstrumentAD} is trained and evaluated on the combination of MedleyDB \cite{bittner2014medleydb} and Mixing Secrets \cite{Gururani2017MixingS}.
Both \cite{Hung2018FramelevelIR} and \cite{Gururani2018InstrumentAD} use frame-level instrument labels for training. In contrast, the model presented by Liu \emph{et al.} \cite{liu2018weakly} uses only clip-level instrument labels associated with YouTube videos for training, using a weakly-supervised approach. 
Both \cite{liu2018weakly} and \cite{Gururani2018InstrumentAD} do not consider pitch information.

As the existing datasets are limited in genre coverage or data size, prediction models trained on these datasets may not generalize well, as shown in \cite{Bittner2017DeepSR} for pitch recognition.
Unlike these prior arts, we explore the possiblity to train a model on large-scale synthesized audio dataset, using a multitask learning method that considers both pitch and timbre.

OpenMIC-2018 \cite{openmic} is a new large-scale dataset for training clip-level instrument recognizers. It contains 20,000 10-second clilps of Creative Commons-licensed music of various genres. But, there is no frame-level labels.


\emph{Multi-pitch streaming} has been referred to as the task that assigns instrument labels to note events \cite{Duan2014MultipitchSO}. Therefore, it goes one step closer to full transcription of musical audio than MPE. However, as the task involves both frame-level pitch and instrument recognition, it is only attempted sporadically in the literature (e.g., \cite{Duan2014MultipitchSO,arora2015multiple}).
By predicting the pianorolls, the proposed model actally performs multi-pitch streaming.


\section{Proposed Dataset}
\label{section_db}
The MuseScore dataset is collected from the online forum of the MuseScore community.
Any user can upload the MIDI and the corresponding audio for the music pieces they create using the software.
The audio is therefore usually synthesized by the MuseScore synthesizer, but the user has the freedom to use other synthesizers.
The audio clips have diverse musical genres and are about two mins long on average. 
More statistics of the dataset can be found from our GitHub repo.

While the collected audio and MIDI pairs are usually well aligned, to ensure the data quality we further run the dynamic time warping (DTW)-based alignment algorithm proposed by Raffel \cite{raffel2016learning} over all the data pairs. We then compute from each MIDI file the groundtruth pianoroll, pitch roll and instrument roll using \texttt{Pypianoroll} \cite{pypianoroll}.

The dataset contains 128 different instrument categories as defined in the MIDI spec. A main limitation is that there is no singing voice.
This can be made up by datasets with labels of vocal activity \cite{kyungyun18ismir}, such as the Jamendo dataset \cite{ramona08icassp}.

Due to copyright issues, we cannot share the dataset itself but the code to collect and process the data.

\begin{figure}[t]
\centering
\includegraphics[width=0.4\textwidth]{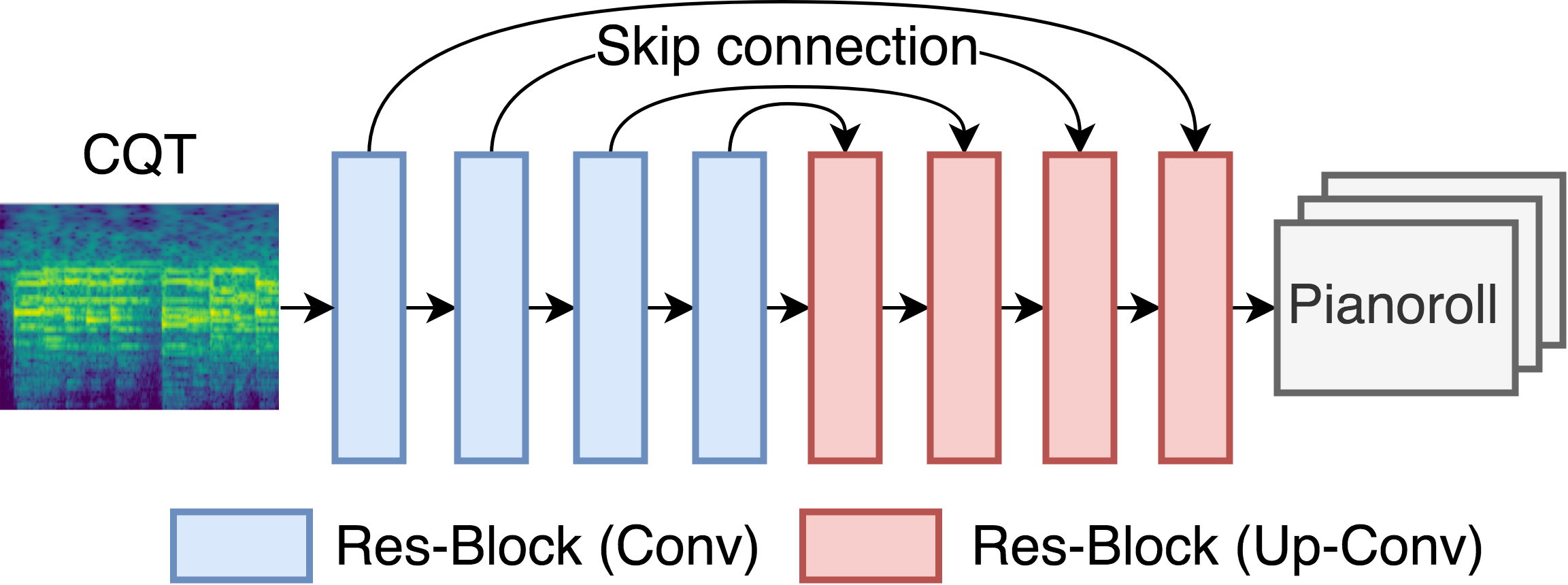}
\caption{The network architecture of the proposed model. It has a simple U-net structure \cite{ronneberger2015u} with four residual convolution layers and four residual up-convolution layers.}
\label{fig: model}
\end{figure}

\section{Proposed  Model}
\label{section_model}

As Fig. \ref{fig: streaming} shows, the proposed model learns a mapping $f(\cdot)$ (i.e., the `Model' block in the figure) between an audio representation $\mathbf{X}$, such as the constant-Q transform (CQT) \cite{schorkhuber2010constant}, and the pianoroll $\mathbf{Y}_{roll} \in \{0,1\}^{F \times T \times M}$, where $F$, $T$ and $M$ denote the number of pitches, time frames and instruments, respectively.
Namely, the model can be viewed as a multi-pitch streaming model. 
The model has two by-products, the pitch roll $\mathbf{Y}_{p} \in \{0,1\}^{F \times T}$ and the instrument roll $\mathbf{Y}_{i} \in \{0,1\}^{M \times T}$.
As Fig. \ref{fig: streaming} shows, from an input audio, our model computes $\widehat{\mathbf{Y}}_{p}$ and $\widehat{\mathbf{Y}}_{i}$ directly from the  pianoroll $\widehat{\mathbf{Y}}_{roll}$ predicted by the model. 
Therefore, $f(\cdot)$ contains all the learnable parameters of the model. 

We train the model $f(\cdot)$ with a multitask learning method by using three cost functions, $L_{roll}$, $L_p$ and $L_i$, as shown in Fig. \ref{fig: streaming}. For each of them, we use the binary cross entropy (BCE) between the groundtruth and the predicted matrices (tensors). The BCE is defined as:
\begin{equation} 
L_* = - \textstyle{\sum} ~[\mathbf{Y}_*\cdot \ln\sigma(\widehat{\mathbf{Y}_*}) + (1-\mathbf{Y}_*) \cdot \ln(1-\sigma(\widehat{\mathbf{Y}_*})) ] \,,
\label{eq:lt}
\end{equation}
where $\sigma$ is the sigmoid function that scales its input to $[0,1]$. 
We weigh the three cost terms so that they have the same range, and use their weighted sum to update $f(\cdot)$. 

In sum, pitch and timbre are modeled jointly with a shared network by our model.
This learning method is designed for music and, to our knowledge, has not been used elsewhere.


\subsection{Network Structure}

The network architecture of our model is shown in Fig. \ref{fig: model}. 
It is a simple convolutional encoder/decoder network with symmetric skip connections between the encoding and decoding layers. 
Such a ``U-net'' structure has been found useful for image segmentation \cite{ronneberger2015u}, where the task is to learn a mapping function between a dense, numeric matrix (i.e., an image) and a sparse, binary matrix (i.e., the segment boundaries). We presume that the U-net structure can work well for predicting the pianorolls, since it also involves learning such a mapping function.
In our implementation, the encoder and decoder are composed of four residual blocks for convolution and up-convolution. 
Each residual block has three convolution, two batchNorm and two leakyReLU layers.
The model is trained with stochastic gradient descent with 0.005 learning rate.
More details can be found from our GitHub repo.

\subsection{Model Input}
\label{section_model_io}

We use CQT \cite{schorkhuber2010constant} to represent the input audio, since it adopts a log frequency scale that better aligns with our perception of pitch. CQT also provides better frequency resolution in the low-frequency part, which helps detect the fundamental frequencies. 
For the convenience of training with mini-batches, each audio clip in the training set is divided into 10-second segments. 
We compute CQT by \texttt{librosa} \cite{mcfee2015librosa}, with 16 kHz sampling rate,  512-sample hop size, and 88 frequency bins. 


\begin{table}
\centering
\begin{tabular}{|r|c c c|} \hline
Method &Instrument &Pitch  &Pianoroll\\
 \hline\hline
$L_{roll}$ only (ablated) &  --- &  --- & 0.623 \\
$L_i$ only (ablated) &  0.896 &  --- & ---\\
$L_p$ only (ablated) &  --- & 0.799 &  ---\\
\hline
all (proposed) & 0.947 & 0.803 & 0.647 \\
\hline
\end{tabular}
\caption{Performance comparison of the proposed multitask learning method (`all') and 3 single-task ablated versions, for frame-level instrument recognition (in F1-score), frame-level pitch recognition (Acc), and pianoroll prediction (Acc) using the triaining and test subsets of MuseScore, for 9 instruments.}
\label{tab:task_comp}
\end{table}

\begin{table*}
\centering
\begin{tabular}{|r| l | c c c c c|c|} \hline
Method& Training set & Piano&Guitar&Violin&Cello&Flute&Avg\\
 \hline\hline
\cite{liu2018weakly} & YouTube-8M \cite{youtube8m} 
&0.766&0.780&0.787&	0.755&0.708&0.759\\
\cite{Gururani2018InstrumentAD} & Training split of `MedleyDB+Mixing Secrets'  \cite{Gururani2018InstrumentAD} 
&0.733&0.783&0.857&0.860&0.851&0.817\\
\cite{Hung2018FramelevelIR}& MuseScore training subset & 
0.690&0.660&0.697&0.774&0.860&	0.736\\
\hline
Ours & MuseScore training subset & 
0.718&0.819&0.682&0.812&	0.961&0.798\\
\hline
\end{tabular}
\caption{AUC scores of per-second instrument recognition on the test split of `MedleyDB+Mixing Secrets', for 5 instruments.}
\label{tab:method_comp}
\end{table*}

\section{Experiment}
\label{section_exp}


\subsection{Ablation Study}

We report two sets of experiments for frame-level instrument recognition. 
In the first experiment, we compare the proposed multitask learning method with its single-task versions, using two non-overlapping subsets of MuseScore as the training and test sets. Specifically, we consider only the 9 most popular instruments\footnote{Piano, acoustic guitar,  electric guitar, trumpet, sax, violin, cello \& flute.} and run a script to pick for each instrument 5,500 clips as the training set and 200 clips as the test set. We consider three ablated versions here: using the U-net architecutre shown in Fig. \ref{fig: streaming} to predict the pianoroll with only $L_{roll}$, to predict directly the instrument roll (i.e. only considering $L_{i}$), and to preidct directly the pitch roll (i.e. only $L_{p}$).

Result shown in Table \ref{tab:task_comp} clearly demonstrates the superiority of the  proposed multitask learning method over the single-task counterparts, especially for instrument prediction. Here, we use \texttt{mir\_eval} \cite{raffel2014mir_eval} to calculate the `pitch' and `pianoroll' accuracies.
For `instrument', we report the F1-score.


\begin{figure}[t]
\centering
\includegraphics[width=0.48\textwidth]{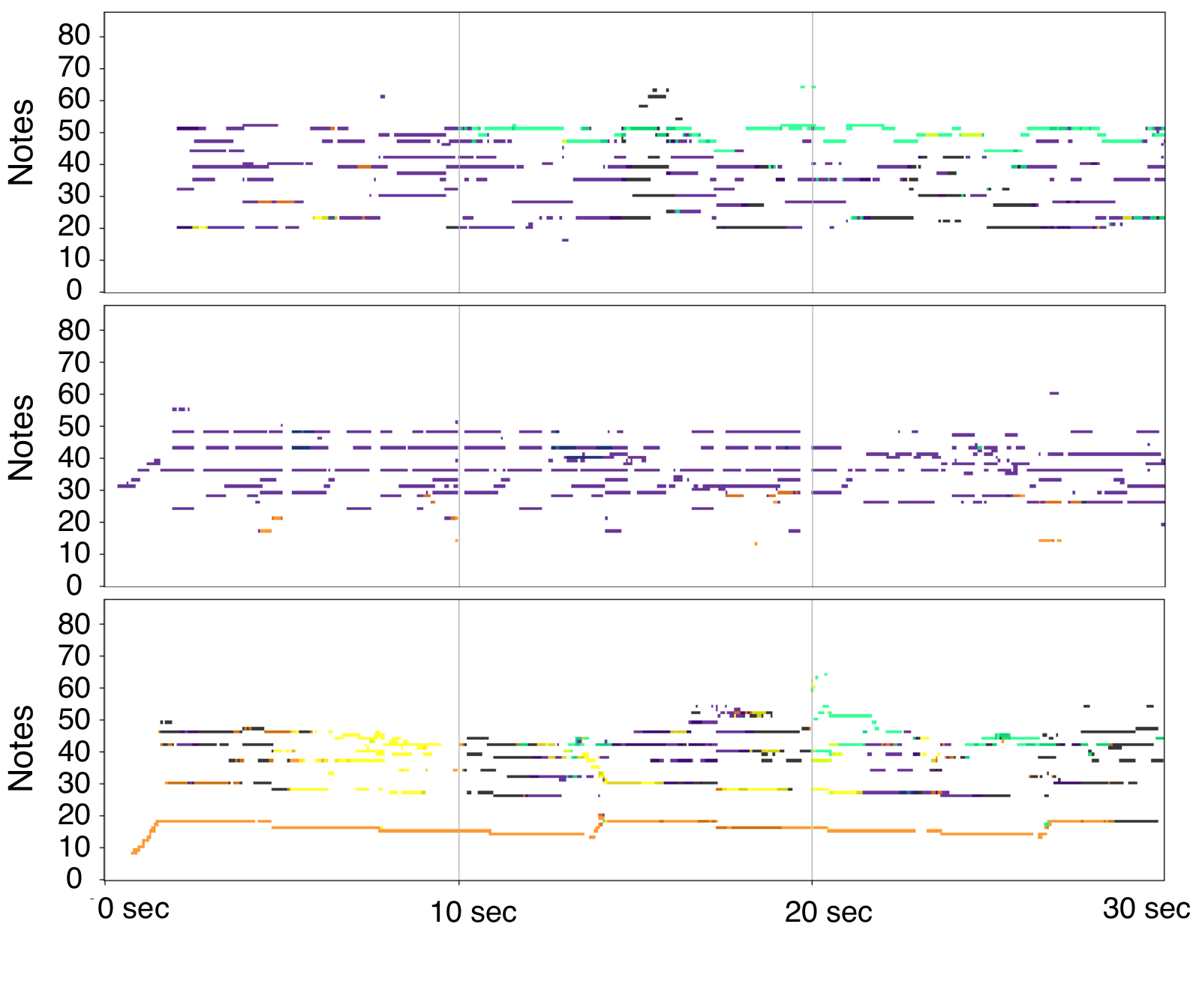}
\vspace{-9mm}
\label{fig: all_of_me}
\caption{The predicted pianoroll (best viewed in color) for the first 30 seconds of three real-world music. We paint different instruments with different colors: \emph{Black}---piano, \emph{Purple}---guitar, \emph{Green}---violin, \emph{Orange}---cello, \emph{Yello}---flute.}
\label{fig: result}
\end{figure}

\subsection{Comparison with Existing Methods}


In the second experiment, we compare our method with three existing methods \cite{liu2018weakly,Hung2018FramelevelIR,Gururani2018InstrumentAD}. 
Following \cite{Gururani2018InstrumentAD}, we take 15 songs from MedleyDB and 54 songs from Mixing Secret as the test set, 
and consider only 5 instruments (see Table \ref{tab:method_comp}). 
The test clips contain instruments (e.g., singing voice) that are beyond these five.
We evaluate the result for per-second instrument recognition in terms of area under the curve (AUC). 

As shown in Table \ref{tab:method_comp}, these methods use different training sets. 
Specifically, we retrain  model \cite{Hung2018FramelevelIR} using the same training subset of MuseScore as the proposed model.
The model \cite{liu2018weakly} is trained on the YouTube-8M dataset \cite{youtube8m}.
The model 
\cite{Gururani2018InstrumentAD} is trained on a training split of `MedleyDB+Mixing Secret', with 100 songs from each of the two datasets. 
The model \cite{Gururani2018InstrumentAD} therefore has some advantages since the training set is close to the test set.
The result of \cite{liu2018weakly} and \cite{Gururani2018InstrumentAD} are from the authors of the respective papers.


Table \ref{tab:method_comp} shows that our model outperforms the two prior arts \cite{Hung2018FramelevelIR,liu2018weakly} and is  behind model \cite{Gururani2018InstrumentAD}. We consider our model compares favorably with \cite{Gururani2018InstrumentAD}, as our training set is quite different from the test set.
Interestingly, our model is better at the flute, while \cite{Gururani2018InstrumentAD} is better at the violin. This might be related to the difference between the real and synthesized sounds for these instruments, but future work is needed to clarify.





\subsection{Multi-pitch Streaming}
Finally, Fig. \ref{fig: result} demonstrates the predicted pianorolls 
for the first 30 seconds of three randomly-selected real-world songs.\footnote{The three songs are, from top to bottom: \emph{All of Me} violin \& guitar cover (https://www.youtube.com/watch?v=YpYQh7eQULc), \emph{Ocean} by Pur-dull (https://www.youtube.com/watch?v=5Lb9GvEO-sA) and \emph{Beautiful} by Christina Aguilera (https://www.youtube.com/watch?v=eAfyFTzZDMM).} 
In general, the proposed model can predict the notes and instruments pretty nicely, especially for the second clip, which contains only a guitar solo. This is promising, since the model is trained with synthetic audio only.
Yet, we also see two limitations of our model. 
First, it cannot deal with sounds that are not  included in the training data---e.g., for the 5th--10th seconds of the third clip, our model mistakes the piano for the flute, possibly because the singer hums in the meanwhile.
Second, it cannot predict the onset times accurately---e.g., the violin melody of the first clip actually plays the same note for several times, but the model mistakes them for long notes. 

\section{Conclusion}
\label{section_conclusion}

In this paper, we have presented a new synthetic dataset and a multitask learning method that models pitch and timbre jointly. It allows the model to predict instrument, pitch and pianorolls representation for each time frame. Experiments show that our model generalizes well to real music.

In the future, we plan to improve the instrument recognition by re-synthesizing the MIDI files from Musescore dataset to produce more realistic instrument sound. Moreover, we also plan to mix the singing voice clips from \cite{bittner2014medleydb} with our training data (for data augmentation) to deal with singing voices.


\bibliographystyle{IEEEbib}
\bibliography{refs}
\end{document}